# Transcranial direct current stimulation to remediate myasthenia gravis symptoms

Ali-Mohammad Kamali [1,2,3,4], Mohammad Reza Hossein Tehrani [3,5], Seyedeh-Saeedeh Yahyavi[1,2,3,4], Siavash Baneshi[3,4], Zahra Kheradmand-Saadi [2,3,6] Masoume Nazeri[5], Maryam Poursadeghfard[5], Mohammad Nami[1,2,3,7]*

[1]Department of Neuroscience, School of Advanced Medical Sciences and Technologies, Shiraz University of Medical Sciences, Shiraz, Iran
[2]DANA Brain Health Institute, Iranian Neuroscience Society-Fars Branch, Shiraz, Iran.
[3]Neuroscience Laboratory, NSL (Brain, Cognition and Behavior), Department of Neuroscience, School of Advanced Medical Sciences and Technologies, Shiraz University of Medical Sciences, Shiraz, Iran
[4]Student research committee, Shiraz University of Medical Sciences, Shiraz, Iran
[5]Clinical neurology research center, Department of neurology, Shiraz University of Medical Sciences, Shiraz, Iran
[6]Department of Foreign Languages and Literature, Shiraz University, Shiraz, Iran
[7]Academy of Health, Senses Cultural Foundation, Sacramento, USA

*Corresponding author*
*Mohammad Nami. Department of Neuroscience, School of Advanced Medical Sciences and Technologies, Shiraz University of Medical Sciences, Shiraz, Iran. torabinami@sums.ac.ir

## Abstract

**Objective:** Myasthenia gravis (MG) is a progressive neurological disease condition characterized by fatigue and muscle weakness. Given the potential untoward effects of current medications used in the management of MG new non-pharmacological approaches including brain stimulation (namely the transcranial direct current stimulation or, in short, tDCS may be considered as potential add-ons to help remediating MG symptoms.

**Methods:** Following a comprehensive neurological and cognitive examinations, ten patients with MG were sequentially enrolled and randomly assigned to either sham or real tDCS delivered over the primary motor cortex for 20 minutes over the first session. In 48 hours, the real arm received sham and the sham arm received real tDCS. After stimulation, the cognitive profile of the patients was evaluated through Cambridge Brain Science-Cognitive Platform. Later, patients' muscular strength (eyes, facial muscle, axial and limb muscles) was examined. In addition, the maximal muscle power was evaluated through the knee extension exercise (1RM) and maximum voluntary contraction (MVC). Likewise, the patients' muscular endurance was recorded through isometric knee extension, sustained hand grip contraction, isometric hip and neck flexion, and isometric shoulder extension indices.

**Results:** Regarding the muscular strength, real vs. sham tDCS improved 1RM and MVC by 18.89% and 15.5%, respectively. Moreover, surface electromyography (sEMG) over the quadriceps femoris muscle amplitude which was recorded during 1RM task was significantly increased after real tDCS by 14.9%. With regard to muscular endurance, the isometric hip and neck flexion task score was increased by 66.5 % and17.89 % in real vs sham tDCS arms, respectively. In addition, anodal stimulation significantly affected isometric knee (by 18.89%) and shoulder (by 36.4%) extension. Meanwhile, the sustained hand grip contraction was not significantly influenced by tDCS.

**Conclusions:** Our findings indicated that brain stimulation exerted no effect on the patients' cognitive functions. The study outcome suggest that tDCS over primary motor cortex may be considered as a potential non-pharmochological treatment add-on in MG. Larger-sized studies need to evaluate the significance of this approach is real-life practice.

1. Introduction

   Myasthenia gravis (MG) is a serious and potentially lethal disease causing diurnal fatigue and progressive muscle weakness. In general, the condition is classified as an autoimmune disorder[1]. Antibodies found in up to 80-90% of patients with MG[2] are known to counteract acetylcholine receptors and eventually destroy them [3]. MG can affect males and females of all age groups with the prevalence rate estimated at about 15 to 179 per one million across studies [4].

   The diagnosis is based on clinical features and laboratory tests. Symptoms include diurnal fatigue fluctuating given the severity of the disease and extent of physical activity [3]. There are 2 types of MG where in one type only ocular muscle involvement exists, while in another, muscle weakness involves bulbar, limb and axial muscles. The extra-ocular muscle weakness is almost asymmetric while limb weakness is symmetric and more prominent in proximal parts [3, 4]. In patients with advanced MG, most muscles including diaphragm might be affected leading to respiratory failure and eventually death. Studies have also shown up to 70-90% decline in acetylcholine receptors due to MG [5]. It has been documented that the thymus gland has a critical role in developing MG through T-helper mediated production of anti-acetylcholine receptor antibodies. In most MG cases, the thymus size is larger than expected causing thymoma in almost 10-25% of patients [5].

   Thymectomy is then recommended especially in patients with early onset disease and without anti-MUSK and anti LRP4 antibodies [3]. Thymectomy can also be considered in patients without thymoma allowing them to achive more favorable treatment response[6].

   In recent comprehensive review, different currently available and future treatment approaches on the basis of MG pathophysiology as well as the auto-antibody and cell response profiles have been outlined [7].

   The quantitative evaluation of treatment response questionnaire in MG (QMG), is perhaps an essential part of therapeutic response prediction based on the quantitative measurement of muscle power and related disease features. Meanwhile, this approach may not necessarily be replaced by clinical assessment and should not be used classify MG patients. This questionnaire may hence be used for evaluation before and during treatment. The MG international committee emphasizes on using all components of questionnaire since the diminished muscle power may occur in one or two domains though showing remission in the overall disease scale score[8].

   The current frequently used medications for MG (e.g.corticosteroids) subject to many untoward effects, therefore, studies have attempted to find alternative solutions owing to less side effects as a replacement or add-ons for such treatments [9, 10]. Among the more recent non-pharmacological treatment options, transcranial direct current stimulation (tDCS) may potentially be considered as a promising approach. The technique is intended to deliver direct electrical current to change brain excitability and consequently the neuronal activity [11]. In fact, factors such as motor learning, muscle strength, fatigue and specific motor skills may be modulated through non-invasive brain stimulation approaches including tDCS [12]. tDCS transmits a weak but stable electrical current (between 1-2 mA) through surface electrodes to the scalp. The duration of intervention typically ranges from 5 to 20 minutes in which the electrical current changes the action potential threshold in neurons[13]. With its efficacy and safety shown in several neurocognitive cognitions, tDCS may potentially be considered as a substitute or add-on for pharmacological treatments in neurocognitive and behavioral conditions [14].

So far, many studies have shown the effectiveness of tDCS for both healthy and patient groups, for instance, it was shown to enhance movement training outcome in CVA patients [15]. Moreover, tDCS resulted in better exercise and movement outcomes through enhancing motivation as well as reducing exercise fatigue[1]. In the present research we hypothesized that tDCS would reduce muscle weakness through bypassing the neuromuscular junction impairment (the core underpinning pathophysiology in MG) hence improving the motor function. As such, might get clinical attention as a promising approach for patients suffering from MG.

## 2. Method
### 2.1. Participants

Thirteen MG patients who were diagnosed through medical criteria and laboratory testing were referred by a neurologist to partake in the study. Three patients were excluded from the study because of severe muscle weakness. The study participants included 1 male and 9 females aging 18 to 50 years with weight between 40 to 100 kg who were under MG treatment for at least 3 months prior to the study. Subjects were included based on not having any associated illness, MG crisis for at least 3 month prior to the study, pregnancy, psychological or neurological disorders, and a history of alcohol or drug use. An informed written consent was obtained from the participants and experimental procedures were approved by the ethical committee at Shiraz University of Medical Sciences (approval No. 1396-01-74-15536). Table 1 outlines the participants' demographic data.

| participants | sex | Age | disease duration | Medication | Thymectomy | Exacerbation | MGQOL | MMT |
|---|---|---|---|---|---|---|---|---|
| 1 | F | 48 | 14 | 1,2 | 2 | 0 | 23 | 13 |
| 2 | F | 38 | 48 | 1,2 | 1 | 2 | 30 | 25 |
| 3 | M | 35 | 9 | 1,3 | 1 | 0 | 6 | 0 |
| 4 | F | 45 | 156 | 1,3 | 1 | 4 | 1 | 3 |

| 5 | F | 39 | 48 | 1 | 1 | 0 | 9 | 17 |
| 6 | F | 30 | 24 | 1,2,3 | 2 | 0 | 17 | 5 |
| 7 | F | 31 | 11 | 1 | 1 | 3 | 7 | 0 |
| 8 | F | 35 | 48 | 1,2,3 | 1 | 0 | 11 | 0 |
| 9 | F | 43 | 36 | 1,2,3 | 1 | 3 | 29 | 11 |
| 10 | F | 51 | 84 | 1 | 2 | 2 | 35 | 4 |

### 2.2. tDCS

A 2mA electrical current for 20 minutes (ramping up and down for 15 seconds) was transferred to the scalp through a two-channel tDCS device (Neurostim-2, Medina Teb). The anode electrodes (35Cm2) were placed over the right (C2, C4, C6, FC2, FC4, FC6) and left (C1, C3, C5, FC1, FC3, FC5) motor cortex (10-20 EEG electrode placement system). The right motor cortex cathode electrode (16 Cm²) was placed over their right shoulder and left motor cortex cathode electrode (16 Cm²) was placed over their left shoulder. Sponges soaked in saline (NaCl 150 mM) were used under the electrodes to transfer the current. In the sham session, after delivering a 30 second current to cause stimulation sensation, the current was switched off.

### 2.3. Instruments and measurements

- **Myasthenia Gravis Activities of Daily Living (MG-ADL):** This questionnaire was used to evaluate the severity of MG systems [16].

- **Myasthenia Gravis Quality of Life 15-Item Scale (MG –QoL15):** This survey was used to asses some quality of life factors of MG patients [17]. The patients were supposed to report how their disease has affected the 15 items in the survey. The survey retains an acceptable construct validity and reliability.

- **Manual Muscle Testing (MMT):** This questionnaire was used to evaluate the maximal power that a muscle could exert [18]

- **Quantitative Myasthenia Gravis questionnaire** (QMG): It includes different sections evaluating double vision, ptosis, swallowing, speech, muscle power of proximal and distal parts of upper limbs and proximal part of lower limbs [16].

- **One-Repetition Maximum (1RM)/ Isometric Knee Extension:** The assessment is an index of the maximal power that a muscle can generate. This index shows the maximum weight that the patients can lift by the knee extension machine. On the other hand, the 1RM or one-repetition maximum index was used to assess the maximum strength measured through 1RM= $w\,(1 + \frac{r}{30})$, considering $r > 1$ (34), where $r$ is the number of repetitions performed and $w$ is the amount of weight. To evaluate isometric knee extension, participants were asked to choose 30% of their 1RM and perform the knee extension exercise once. The time that they could hold the lift was then recorded.

- **Maximum Voluntary Contraction (MVC)/ Sustained Hand Grip Contraction**: The assessment tests the muscle power in patients suffering from neuromuscular disease. MVC was assessed by a hand-grip after both sham and real tDCS sessions. To assess the sustained hand-grip contraction, subjects were asked to choose 30% of their MVC and perform the hand contraction by a dynamometer (SAEHAN, SH1003). The duration they kept gripping the dynamometer with the specified weight was recorded.

-**The Cambridge Brain Science-Cognitive Platform (CBS-CP):** The overall cognitive profile of the patients was recorded through CBS-CP ( an online platform addressing cognitive abilities) after brain stimulation. Three tasks were chosen among the three higher-order cognitive components i.e. reasoning, memory and verbal ability [19, 20]. Based on the consultation received from a panel of three cognitive scientists, the Odd One Out, Paired Association and Digit Span tasks were selected from the reasoning, memory, and verbal domains, respectively. Moreover, the average scores in each task was compared to the mean scores of each task (z-score) within the CBS database.

- **Head lift and leg outstretch:** The patients were asked to sequentially bend their right and left legs at a 45-degree angle. The average time of holding the bent left and right legs was recorded as the isometric hip flexion index. Moreover, the duration over which patients could hold their neck at a 45-degree angle was considered as the isometric neck flexion index.

- **Arm Outstretch:** the amount of time patients could extend their hands at a 90-degree angle was recorded by a chronometer and considered as their isometric shoulder extension index.

- **Surface Electromyography (sEMG):** SEMG was recorded from the rectus femoris muscle by a NeXus Biofeedback setup (NeXus 10 MKII, Mind Media). The sensors were attached to the midpoint of anterior superior iliac spine and patella through chest leads and the sEMG was recorded during the 1RM task.

## 2.4. Data analysis
To compare the differences between sham and real tDCS (in terms of MVC, sustained hand-grip contraction, 1RM, isometric knee extension, sEMG, isometric shoulder extension, isometric neck and hip flexion, and CBS-CP), a series of paired sample *t*-tests were done. The Mean±SEM (Standard Error of Mean) was considered to report any statistical significance in outcomes following the sham and real tDCS sessions. The statistically significant *p* values were set at 0.05. The SPSS statistical package (Version 25.0.0, Copyright©IBM) was used for data analysis.

## 2.3. Experimental design
At the beginning of the study, every participant was asked to complete the MG-ADL and MG-QOL15 to get his/her personal information and disease characteristics recorded.

The MMT was also used to obtain qualitative evaluations of muscle power before and after both sham and real tDCS. All examinations were done by a same neurologist over 2 sessions as per the instructions in QMG manual [8]. The entire set of questionnaires used in the present investigation was originally recommended for clinical research by the Task Force of the Medical Scientific Advisory Board- the Myasthenia Gravis Foundation of America [21]. Through a single-blind, counter-balanced design, data were obtained over two sessions over a 48-hour interval. Through consecutive randomization, patients were randomly assigned to sham and real tDCS in order to eliminate the learning and practice effects. Patients were randomly assigned to 2 mA sham or real tDCS over the motor cortex for 20 minutes in the first session. After 48 hours, the group which received sham first, received real tDCS and vice versa. After the brain stimulation, to record the patients' cognitive profiles, subjects were required to perform 3 tasks from the CBS-CP components of reasoning, memory and verbal abilities. Patients were then asked to hold the hand-grip dynamometer at a 90-degree angle for 3 times with the maximum weight they could bear. The average score of the three-time exercises was considered as their MVC. Then, to evaluate the patients' sustained hand-grip contraction, patients were asked to choose 30% of their MVC and perform the hand contraction by a dynamometer. The amount of time that they could bear the dynamometer with the specified weight was recorded by a chronometer.

Similarly, the patients were required to perform the knee extension exercise at a 90-degree angle for at least 6 to 12 times with the maximum weight they could bear with the Knee Extension Machine in order to obtain their 1RM. Later, to measure the isometric knee extension, patients were asked to choose 30% of their 1RM and hold the lift through the knee extension exercise. The duration they could bear the lift was recorded by a chronometer. Furthermore, sEMG was recorded during the 1RM exercise. Moreover, in order to measure the patients' isometric hip flexion, the duration that they could hold their leg at a 45-degree angle was recorded by a chronometer. Likewise, the amount of time they could hold their neck at a 45-degree angle was recorded as the patients' isometric neck flexion. At the end of the experiment, the patients' arm out stretch indices were also recorded.

### 3. Results
### - 1RM/ Isometric Knee Extension
With regard to the 1RM used to evaluate the maximal muscular power, the real tDCS *vs.* sham tDCS could increase the muscular strength mean scores by 18.89% ($p$=0.03). Moreover, real tDCS was found to be effective in improving the isometric knee extension time by 37.62% ($p$=0.003) (Figure 1).

- **MVC/ Sustained Hand Grip Contraction**
Based on our findings, compared to sham stimulation, the anodal stimulation over the motor cortex resulted in a significant increase of the MVC average score by 15.5% ($p$=.003). Meanwhile, the hand-grip contraction time did not show any statistically significant change in sham *vs.* real tDCS session (Figure 1).

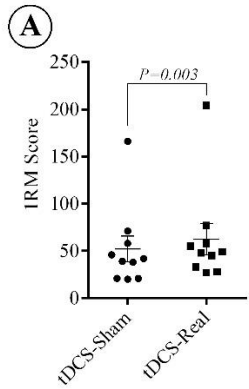
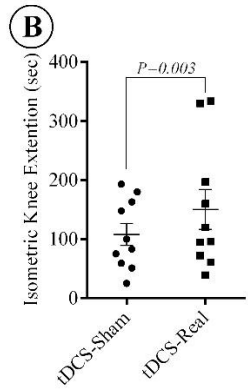
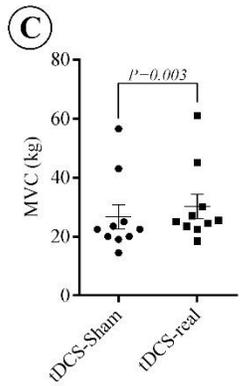
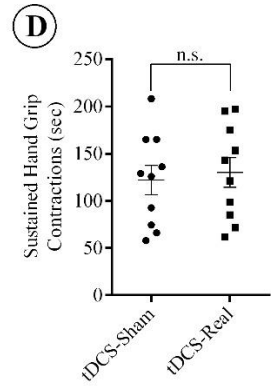
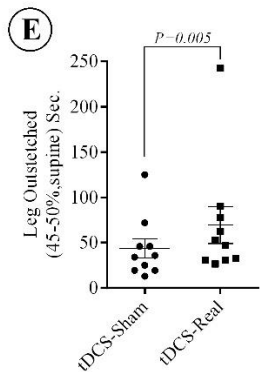
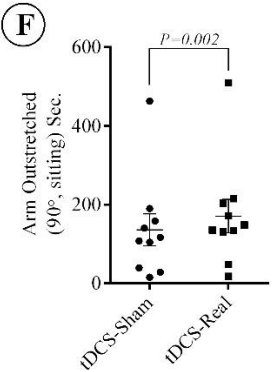
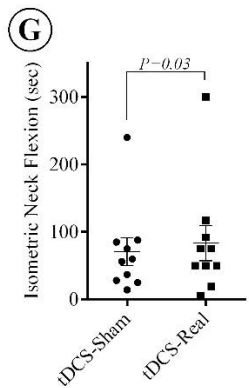

- **Head Lift and Leg outstretch**

Considering the patients' lumbar and cervical muscular power, it was shown that the patients' leg outstretch time could increase by 66.5 %( *p*=0.005) following the anodal motor cortex stimulation. Similarly, the isometric neck flexion time improved by 17.89 %( *p*=0.03) after real tDCS (Figure 1).

- **Arm Outstretch**

Comparing the isometric shoulder extension time after sham and real tDCS, findings revealed that the anodal stimulation could significantly increase the outcome by 36.4%(*p*=0.002) (Figure 1).

- **QMG**

Considering the patient's quantitative myasthenia gravis testing form sore, it was shown that the patients' QMG score could increase by 20.8 %( *p*=0.01) following the anodal stimulation over motor cortex (Figure 2).

- **sEMG**

The findings revealed that the sEMG frequency and amplitude recorded during 1RM significantly increased after real *vs.* sham tDCS by 14.9%(*p*=0.02) (Figure 3).

- **CBS-CP**

In terms of the patients' cognitive performance assessed through CBS-CP , results of the paired-sample *t*-tests showed no statically significant difference between the sham and real tDCS sessions in terms of memory, reasoning and verbal tasks (Figure 4).

However, the average memory ($p=.001$) and verbal ($p=0.01$) z-scores in the sham session were significantly lower than the mean scores in CBS database (Figure 5).

**Figure 1.** Dot plots show the MG patients' muscular strength consisting 1RM and MVC and muscular endurance including Isometric hip and neck flexion, isometric knee and shoulder extension, and sustained hand grip contraction. Panel (a) indicates a significant difference between the 1RM of the MG patients in sham and real tDCS sessions ($p <0.05$). 1RM is obtained from 1RM= $w \left(1 + \frac{r}{30}\right)$ ($r >1$ and is the number of repetitions performed and $w$ is the amount of weight). Panel (b) shows a significant difference between the MG patients' isometric knee extension time in sham and real sessions ($p<0.05$). Panel (c) indicates a significant difference in MVC (maximum voluntary contraction, the average of holding the hand-grip dynamometer at a 90-degree angle with maximum power for 3 times) from sham to real tDCS session ($p<0.05$). Panel (d) shows no significant difference between the patients' sustained hand-grip contraction time in sessions 1 and 2. Panel (e) shows a significant difference between isometric knee flexion time in sessions 1 and 2 ($p<0.05$). Panel (f) shows a significant difference between isometric hip flexion time in sham and real tDCS sessions ($p<0.05$). Panel (g) indicates a significant difference between isometric knee flexion time in sham and real tDCS sessions ($p<0.05$). Paired $t$-test was used with the $p$ value at .05. n. s., nonsignificant.

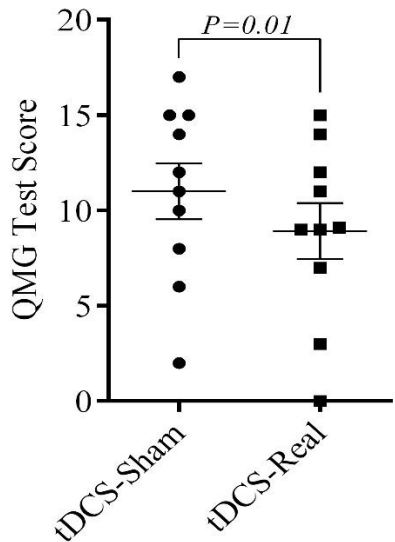

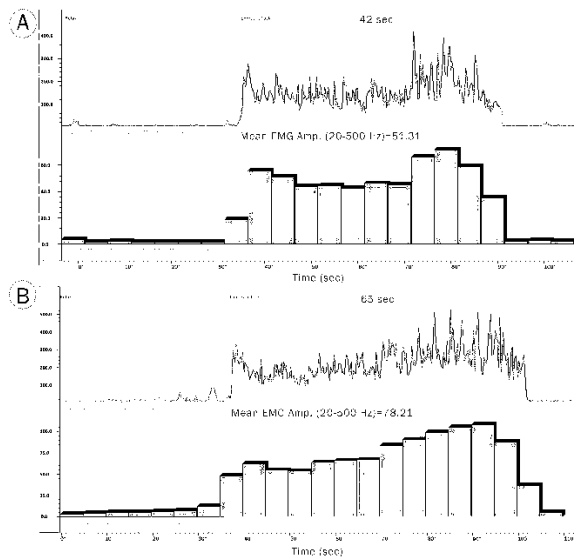

**Figure 2.** It was shown that the patients' QMG score could increase ($p=0.01$) following the anodal stimulation over motor cortex

**Figure 3.** This figure represents sEMG of MG patients during 1RM exercise. Panels (a) and (b) shows sEMG frequency after sham and real tDCS, respectively. The peaks represent lifts during 1RM exercise. Comparing panels (a) and (b), real tDCS increased the sEMG frequency. Paired $t$-test was used with the $p$ value at .05. n. s., nonsignificant

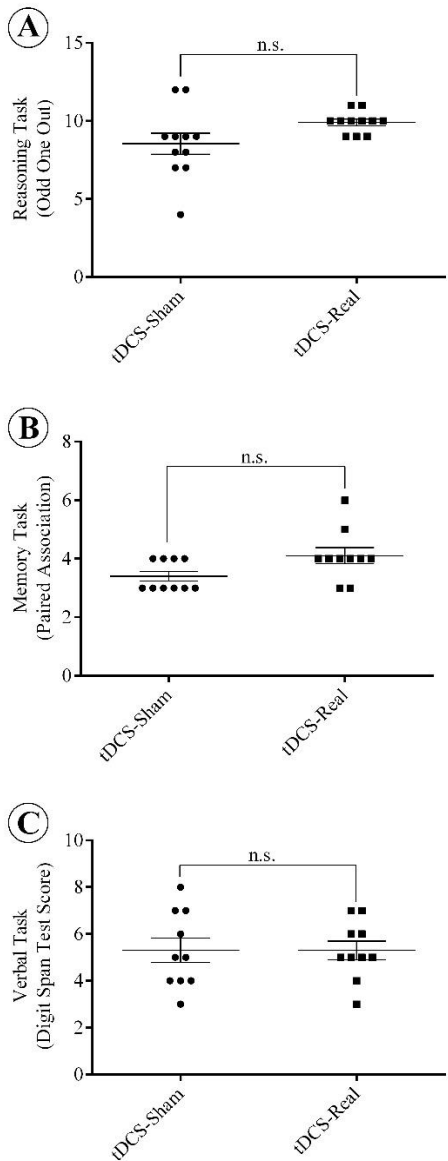

## 4. Discussion

The present findings revealed that tDCS can improve MG symptoms, a disease which is characterized by fatigue and muscle weakness with the worldwide prevalence of 40-180 million people [22]. MG is an autoimmune disease in which antibodies are produced against acetylcholine receptors in the blood and destroy the nicotinic acetylcholine receptors causing them to bind to each other, resulting in endocytosis of the acetylcholine receptors in active synapses [23, 24]. Nearly, all patients with MG require treatment [25]. The first line of treatment for MG include immune-modulators and acetylcholinesterase inhibitors [26] which subject to several side effects. So far, no study has examined the possible beneficial effects of motor cortex stimulation in such patients. The majority of MG studies have focused on immune system whereas studies on neural aspects in MG are scarce. The main research question was to examine how much the primary motor cortex stimulation may improve motor functions in such patients. tDCS is a safe brain stimulation method [27]; whereas, the majority of effective pharmacotherapies are associated with certain complications. Further studies on tDCS may potentially introduce it as a complementary method to reduce MG symptoms possibly with lower required doses of acetylcholinesterase inhibitors. tDCS can reduce the stimulation threshold of the large pyramidal neurons [28]. It is considered as an effective approach in improving major depression [29] and a promising treatment for generalized anxiety disorder with established effectiveness [30]. tDCS may even be used to improve the neurocognitive functionality inhealthy people [31].

Figure 4. Dot plots representing each MG patient's performance on CBS-CP tasks of reasoning, memory, and verbal abilities after sham and real tDCS with the interval of 48 hours. Panel (a) shows no significant difference between the patients' performance on a memory task in sham and real tDCS sessions. Panel (b) indicates no significant difference between the patients' performance on a reasoning task in sham and real sessions. Panel (c) represents no significant difference in a memory task scores from sham to real tDCS session. Paired *t*-test was used with the *p* value at .05. n. s., nonsignificant. CBC-CP: Cambridge Brain Science-Cognitive Platform

MG cannot be effectively managed in 10-15% of the patients who continue to suffer from severe immunosuppressive complications [32]. Meanwhile, alternative non-pharmacological methods can be considered for such patients. Proximal muscle weakness is among the important symptoms of MG [33]. In this study, a single session of anodic stimulation over motor cortex notably reduced the patients' weakness. The MG-induced muscle weakness after a period of rest or following the administration of acetylcholinesterase inhibitors, such as neostigmine or pyridostigmine, is relatively remarkable [26]. This suggests that increased acetylcholine concentration in the synaptic space can play an important role in reducing the symptoms of the disease through inhibition of acetylcholinesterase [34]. In this study, stimulation of the motor cortex could plausibly result in more acetylcholine release. This hypothesis would however require further research to get tested. Further studies are needed to determine the extent to which more brain stimulation sessions can further improve disease symptoms. Patients who participated in the study continued to take their medicines prescribed before the study. More controlled studies would be needed to show possibility of reducing or discontinuing acetylcholinesterase inhibitors upon co-administration of tDCS. The disease was more prevalent in women than men in the current study, which is in line with previous studies [35].

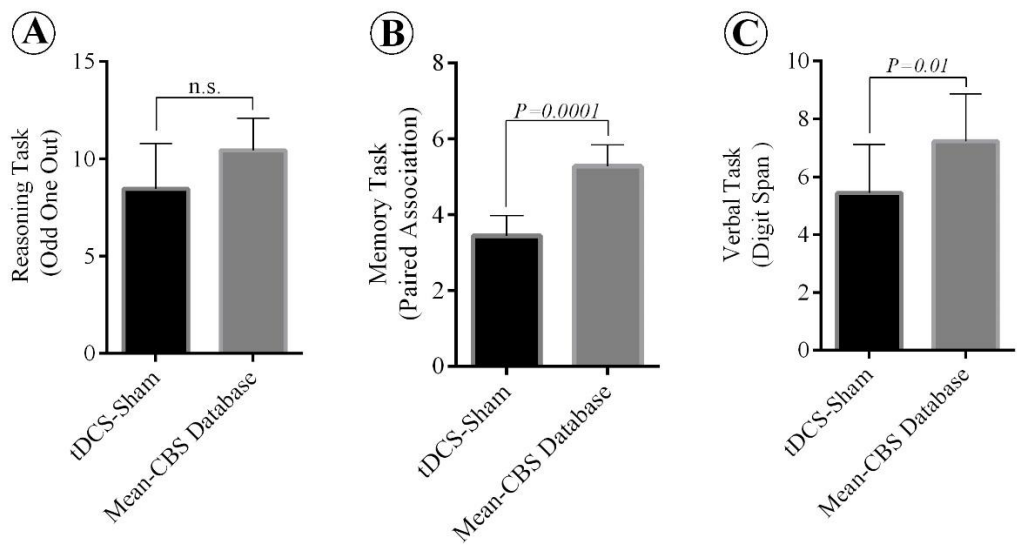

**Figure 5.** Bar graphs representing MG patients' average performance on CBS-CP tasks of reasoning (Odd One Out), memory (Paired Association), and verbal (Digit Span) abilities after sham tDCS and the mean CBS database on these three tasks assessed in this study. Panel (a) shows a significant difference between the patients' performance on a memory task in the sham session and the mean CBS database. Panel (b) indicates lower performance of MG patients on a verbal task in the sham session compared to the mean CBS database. Panel (c) represents no significant difference between the MG patients in the sham session and mean CBS database on a reasoning task. Paired t-test was used with the p value at .05. n. s., nonsignificant. CBC-CP: Cambridge Brain Science-Cognitive Platform

tDCS is a non-invasive treatment probably with fewer side effects than pharmacotherapy. It may be possibly considered as effective in improving the symptoms of MG in patients who are medication refractory or intolerant. Moreover, it may also be a proper therapeutic solution for patients who do not want to take medication. However, our study was subject to some shortcomings including the sample size. The lower number of male subjects owing to the higher prevalence of MG among women, resulted in a smaller statistical population. Although MG has different subgroups [36], this study investigated it as a whole. Furthermore, patients' brain mapping through quantitative electroencephalography (qEEG) while performing different cognitive and functional tasks would provide information on neurodynamics in future studies.

## 5. Conclusion

Taken together, our findings suggest that the anodal stimulation of the primary motor cortex vs. sham may improve MG symptoms and enhance patients' muscular strength. In other words, while tDCS was not effective in improving patients' cognitive abilities, it may be a promising technique for physical empowerment of MG patients. The present results may pave the path for neurologists and neuroscientists to establish modern techniques for treating MG patients through non-pharmacological methods.

## Acknowledgments
The study was financially supported by Dana Brain Health Institute; Iranian Neuroscience Society, Fars Chapter, Shiraz, Iran. Additionally, Shiraz University of Medical Sciences funded the study under the grant No. 1396-01-74-15536.

## Authors' contributions
The authors declare that the research was conducted in the absence of any commercial or financial relationships that could be construed as a potential conflict of interest. All authors read and approved the final manuscript.

### *Study funding*
This work was supported by Shiraz University of Medical Sciences under the grant No. 1396-01-74-15536.